# Development of Ferroelectric Order in Relaxor (1-x)Pb(Mg$_{1/3}$Nb$_{2/3}$)O$_3$ - xPbTiO$_3$ ( 0≤ x ≤0.15)

Z.-G. Ye,[*]  Y. Bing,  J. Gao, and A. A. Bokov
*Department of Chemistry, Simon Fraser University,
Burnaby, BC, V5A 1S6, Canada*

P. Stephens
*Department of Physics and Astronomy, State University of New York
at Stony Brook, New York 11794-3800*

B. Noheda  and  G. Shirane
*Physics Department, Brookhaven National Laboratory,
Upton, New York 11973 -500*

(Dated: August 15, 2002)

The microstructure and phase transition in relaxor ferroelectric Pb(Mg$_{1/3}$Nb$_{2/3}$)O$_3$ (PMN) and its solid solution with PbTiO$_3$ (PT), PMN-xPT, remain to be one of the most puzzling issues of solid state science. In the present work we have investigated the evolution of the phase symmetry in PMN-xPT ceramics as a function of temperature (20 K < $T$ < 500 K) and composition (0 ≤ $x$ ≤ 0.15) by means of high-resolution synchrotron x-ray diffraction. Structural analysis based on the experimental data reveals that the substitution of Ti$^{4+}$ for the complex B-site (Mg$_{1/3}$Nb$_{2/3}$)$^{4+}$ ions results in the development of a clean rhombohedral phase at a PT-concentration as low as 5%. The results provide some new insight into the development of the ferroelectric order in PMN-PT, which has been discussed in light of the kinetics of polar nanoregions and the physical models of the relaxor ferroelectrics to illustrate the structural evolution from a relaxor to a ferroelectric state.

## I. INTRODUCTION

Research on relaxor ferroelectrics and related materials has undergone an accelerated growth both in fundamental understanding of the structural and physical properties and in practical applications. The complex perovskite Pb(Mg$_{1/3}$Nb$_{2/3}$)O$_3$ [PMN] exhibits typical relaxor ferroelectric properties,[1] characterized by the frustration of local polarization due to composition and field inhomogeneities, which prevents long-range ferroelectric order from developing completely. The mechanism of the relaxor ferroelectric behavior continues to be a fascinating puzzle. Experimental data from neutron scattering,[2] nuclear magnetic resonance (NMR)[3] and measurements of nonlinear dielectric susceptibility,[3,4] pointed to the existence of a nonergodic structural glassy state in relaxors below a certain temperature. It is usually believed[3,5,6] that reorientational polar species responsible for the glassy behavior are composed of the clusters of low-symmetry structure having a size of a few nanometers, which were observed in relaxors. Recently, by means of dielectric spectroscopy, Bokov and Ye[7-9] have discovered a universal relaxor dispersion in PMN and related materials with different degrees of ferroelectric ordering, and showed that it is an important common property of relaxor ferroelectrics. The universal relaxor polarization is described by a microscopic model consisting of 'soft' polar nanoregions (PNRs) with unit cells that can freely choose several different directions, while the direction of the total moment of the nanoregion remains the same.[9] Such approach makes it possible to apply a standard spherical model to relaxor ferroelectrics, which predicts the experimentally observed quadratic divergence of susceptibility above the critical temperature. This model is complementary to the so-called spherical random bond – random field (RBRF) model proposed to explain the NMR data and the non-linearity of the total dielectric susceptibility in relaxors.[3,6]

From the structural point of view, the average symmetry of PMN, when probed by conventional x-ray or neutron diffraction techniques, appears to be cubic ($Pm\bar{3}m$) down to very low temperature with no evidence of macroscopic structural phase transition taking place through or below the temperature of maximum permittivity ($T_{max}$ = 265 K, at 1 kHz). However, diffuse scattering on the diffraction patterns were observed below the so-called Burns temperature $T_d$ (or $T_B$) ≈ 600 K, indicating the presence of locally polar regions of rhombohedral *R3m* symmetry, the volume ratio of which (over the cubic matrix) increases upon cooling to reach about 25% at 5 K.[10-13] Such an evolution of local structure is in agreement with the deviations from linearity of the refractive index,[14] the lattice parameters, the thermal expansion and strain,[10,15] which result from the presence PNRs. The dielectric permittivity shows a deviation from the Curie-Weiss law below $T_d$. A macroscopic ferroelectric phase can be induced by application of an electric field along $[111]_{cub}$, associated with cubic to rhombohedral symmetry breaking, as revealed by optical domain, dielectric, polarization and structural studies.[16,17] Recent neutron scattering studies have identified a ferroelectric soft mode in PMN at 1100 K that becomes overdamped near 620 K, suggesting a direct connection between the soft mode and the PNRs.[18] More interestingly, at lower temperature the soft mode in PMN reappears close to $T_C$ = 213 K,[19] temperature at which the electric field-induced polarization vanishes spontaneously upon zero field heating,[16] and a sharp peak in the temperature dependence of the hypersonic damping appears.[20] To interpret the measured intensities of the diffuse scattering in PMN in accordance with the concept of ferroelectric soft mode, Hirota *et al.*[21] has proposed, and demonstrated the validity of, a phase-shifted condensed soft mode model of the PNRs, i. e., the displacement of PNRs along their polar directions relative to the surrounding cubic matrix (H-shift). Therefore, the phonon dynamics clearly indicates the ferroelectric nature of the relaxor PMN below $T_C$, although the average structure of the system remains cubic.

---

[*] Corresponding author, email: zye@sfu.ca

The substitution of $Ti^{4+}$ for the complex $(Mg_{1/3}Nb_{2/3})^{4+}$ ions in the solid solution system between relaxor PMN and ferroelectric PbTiO$_3$ [PT], (1-x)PMN-xPT [PMN-xPT], results in a macroscopic ferroelectric phase with a morphotropic phase boundary (MPB) initially located at x ≈ 0.33,[22-24] separating the rhombohedral (relaxor side of the solid solution) and tetragonal (ferroelectric side) phases. Single crystals of PMN-PT with composition near the MPB are technologically promising materials for electromechanical transducers thanks to their excellent piezoelectric performance.[25] Intensive investigation has recently been undertaken on the structure of the MPB phases and the related properties.[26] A new MPB phase diagram of the PMN-PT system has recently been established by Noheda *et al.* based on high resolution synchrotron x-ray diffraction data, which reveals the presence of an intermediate monoclinic phase existing in between the rhombohedral and the tetragonal phases.[27] Therefore, a spectrum of structures and a variety of macroscopic properties can be tuned by $Ti^{4+}$ substitution, ranging from disordered relaxor state, to a ferroelectric rhombohedral *R3m*, monoclinic *Pm* ($M_C$) and tetragonal *P4mm* phase.

The PMN-xPT solid solutions with low concentration of PT (x ≤ 0.15) are also important materials for a wide range of applications, such as multilayered capacitors, electromechanical transducers, etc., especially in the form of ceramics.[28] However, different results have hitherto been reported by different groups on the structural nature of those compositions. Jang *et al.*[29] found that PMN-xPT with x ≤ 0.13 exhibits typical relaxor ferroelectric behavior associated with a PMN-like pseudocubic structure. X-ray structural studies on x=0.05 and 0.10 single crystals by Bunina *et al.* seemed to confirm that result.[30] Based on neutron powder diffraction data, Fujishiro *et al.* reported that no long-range order can develop in PMN or PMN-0.1PT at low temperature unless an electric field is applied which induces a macro polar phase.[13] Recently, Dkhil *et al.*[31] observed that in PMN the intensity of the diffuse scattering on the x-ray and neutron diffraction patterns appearing at about 350 K increases upon cooling and plateaus below $T_f$ ≈ 200 K, at which an abrupt change occurs in the slop of the variation of the lattice constant. Such an anomaly provides a supplementary evidence for the ferroelectric ordering in unpoled PMN, consistent with the phonon energy recovery[19] and the hypersonic damping anomaly at $T_C$ = 213 K[20] (Note that the temperature $T_f$ identified in Ref. 31 actually corresponds to the $T_C$ of PMN discussed above). In PMN-0.10PT, the diffuse intensity was found to increase upon cooling and shows an unusual critical divergence at $T_C$ = 285 K (although the *q*-value was not specified), associated with a structural transition into a long-range rhombohedral phase.[31] On the other hand, Koo *et al.* showed that, in PMN-0.20PT, the diffuse intensity on neutron scattering at *Q* = (1.04, 0, 0.96) peaks near $T_C$ ≈ 380 K, but stays up below $T_C$ without divergence.[32]

Therefore, such fundamental questions as what the nature of ferroelectric ordering in PMN is and how the PNRs develop into a mesoscopic or macroscopic ferroelectric state upon substitution of PT in the PMN-xPT solid solution, still remain unclear. The objective of the present work is to investigate the variation of the phase symmetry in PMN-xPT ceramics (x ≤ 0.15) as a function of temperature and composition. The structural analysis based on the experimental data from high-resolution synchrotron x-ray diffraction allows us to reveal the establishment of a rhombohedral polar phase in PMN-xPT with x ≥ 0.05. The results are discussed in light of the kinetics of polar nanoregions and the physical models of the relaxor ferroelectrics to illustrate the structural evolution from a relaxor to a ferroelectric state.

## II. EXPERIMENT

The complex perovskite samples of (1-x)Pb(Mg$_{1/3}$Nb$_{2/3}$)O$_3$ – xPbTiO$_3$ [PMN-xPT or PMN-100xPT], with x = 0, 0.05, 0.10, 0.15 and 0.30, were prepared in the form of ceramics using an improved two-step columbite precursor method.[33] In the first step, the starting oxides of MgO and Nb$_2$O$_5$ (> 99.9%) were thoroughly ground in ethanol, cold-pressed into a pellet and calcined at 1100 $^o$C for 12h to form the pure columbite phase of MgNb$_2$O$_6$ (The stoichiometry was achieved by careful weight compensation of MgO based on our thermal gravimetric analysis, since the commercial MgO is well known to adsorb a significant amount of various gas molecules from the atmosphere). In the second step, the columbite precursor powder was reacted with PbO or PbO and TiO$_2$, according to PbO + xTiO$_2$ + [(1-x)/3]MgNb$_2$O$_6$ → Pb[(Mg$_{1/3}$Nb$_{2/3}$)$_{(1-x)}$Ti$_x$]O$_3$, with 2wt% excess of PbO added to compensate for the evaporation of PbO during the subsequent calcining and sintering processes. Each composition was thoroughly grounded and calcined at 900 $^o$C for 4h. The calcined powder was then reground with the addition of a few drops of binding agent (polyvinyl alcohol, PVA) and cold-pressed into a pellet of about 3mm thick and 15mm in diameter, which was first heated up to 650 $^o$C in a Pt crucible for 1h in open air to drive off the PVA, and then sintered at 1230 $^o$C for 4h in a sealed alumina crucible with PbO-enriched atmosphere to form high-density ceramics. The light yellow ceramic pellets were polished with fine diamond paste and ultrasonically cleaned. X-ray patterns obtained on a conventional diffractometer with Cu Kα confirmed the formation of pure perovskite phase with no evidence of any impurities.

Synchrotron x-ray powder diffraction experiments were performed on unpoled samples at beamlines X7A (for PMN-5PT and PMN-10PT) and X3B1 (for PMN and PMN-15PT) at the Brookhaven National Synchrotron Light Source. A double-crystal channel-cut Si (111) monochromator was used in combination with a Ge (220) analyzer and a scintillating detector. The wavelength was set to λ ≈ 0.7 Å and calibrated with a Si reference standard. The resulting instrumental resolution (full width at half maximum, FWHM) is about 0.005° on the 2θ scale, an order of magnitude better than that of a conventional laboratory instrument. At X7A, the data were collected directly from the ceramic pellets loaded on a symmetric "flat-plate" reflection (Bragg-Brentano) geometry with step-scans at 0.005° intervals over selected angular regions while the sample was rocking over 1 – 2° to obtain better powder averaging results. The measurements as a function of temperature were performed between 20 and 500 K. For low temperature measurements, the samples were loaded in a closed-cycle He cryostat, which has an accuracy of about ±2 K. The measurements above room temperature were carried out with the samples loaded inside a wire-wound BN tube furnace with an accuracy of ±3 K. At X3B1, the data were collected with rotating capillaries and Si powder was used as a standard at each temperature so that the accuracy in the calculated lattice parameters was within 0.0003 Å. In most of the cases, the scans were carried out over narrow angular regions centred about the six pseudocubic reflections (100), (110), (111), (200), (220) and (222), in order to determine the crystal symmetry and related lattice parameters. For data analysis, the individual reflection profiles were fitted by least-squares to a pseudo-Voigt function appropriately corrected for asymmetry,[34] with intensity, peak position and peak-width (FWHM) as variables, to determine the lattice parameters.

## III. RESULTS

The selected (222) or (111) diffraction patterns of PMN, PMN-5PT and PMN-15PT at low and high temperature (below and above $T_C$) are given in Figure 1. It can be seen that at high temperatures, all the three samples exhibit a single (222) or (111) reflection peak, in agreement with the perovskite cubic symmetry of the PMN-PT system. At 50 K, pure PMN basically shows a single (222) peak except for a weak diffuse kink that occurs at 2θ = 34.84°. The (111) reflection of PMN-5PT no longer remains singlet at 15 K, but splits with a shoulder that peaks at a lower angle. The (111) reflection of PMN-10PT exhibits a clear split at low temperatures with a narrow line width (not shown). The lower row in Fig. 1 shows the (222) reflection of PMN-15PT that clearly splits into two peaks located at 2θ = 34.87° and 35.02° with an intensity ratio of about 1:2, as expected for a pure rhombohedral phase. The splitting of the (111) or

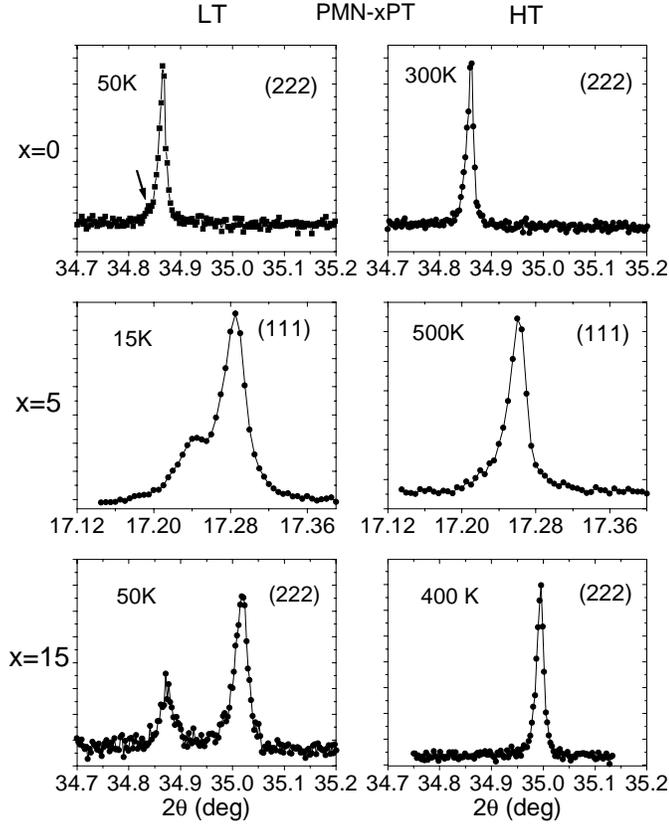
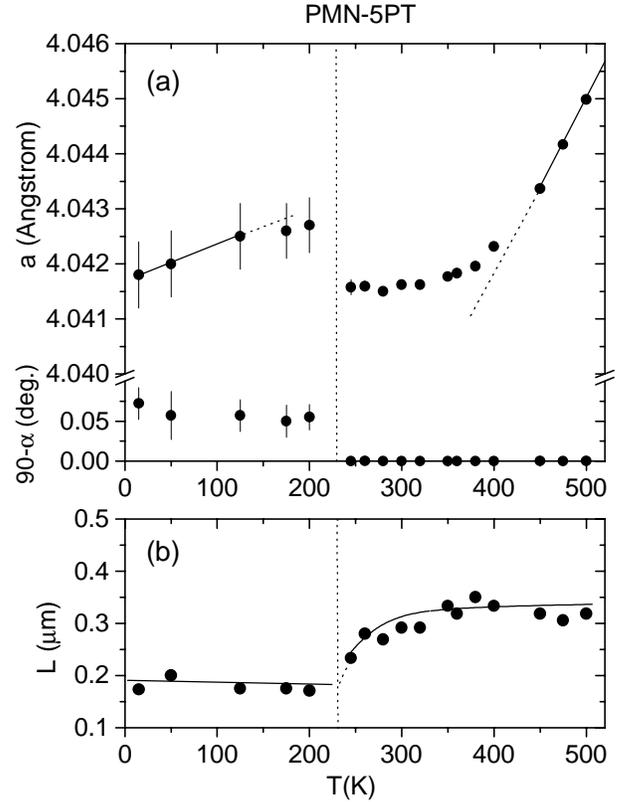

FIG. 1 Selected (222) or (111) diffraction patterns of PMN-xPT (x = 0, 5 and 15%) ceramics at low and high (above $T_C$) temperatures.

FIG. 2 Variations of (a) the cubic or rhombohedral lattice constants, $a$ and $\alpha$, and (b) the coherence length ($L$), as a function of temperature for PMN-5PT.

(222) reflection in PMN-xPT ($0.05 \leq x \leq 0.15$) indicates the formation of a rhombohedral phase at low temperatures. Note that the data for PMN-5PT and PMN-10PT were measured on the ceramic pellets, and in principle, are not free from the effects of preferred orientation and texture that affect the intensity profiles (but do not alter the lattice parameters so determined). That is why the intensity ratio of the (111) peaks for PMN-5PT looks more like 1:3, rather than 1:2, as observed for PMN-15PT in the same figure.

Figure 2(a) presents the variation of the cubic or rhombohedral lattice parameters, $a$ and $\alpha$, as a function of temperature for PMN-5PT. Upon cooling, the unit cell dimension $a$ first decreases linearly until about 430 K, then considerably curves up with a nearly zero slope between 35 and 245 K. Then, it undergoes an abrupt increase between 245 and 200 K, which is accompanied by a slight deviation from 90° of the angle $\alpha$ ($\Delta\alpha \approx 0.05°$), indicating the onset of the rhombohedral phase. Upon further cooling, $a$ decreases almost linearly below 125 K while $\alpha$ increases slightly. In Figure 2(b), the variation of the coherence length ($L$) is shown as a function of temperature. L has been extracted from the Williamson-Hall plots[35]: $\Gamma\cos\theta = \lambda/L + 2(\Delta d/d)\sin\theta$, where $\Gamma$ is of the FWHM. The $\Gamma\cos\theta$ vs. $\sin\theta$ relation was plotted for each temperature from the (100), (110), (111), (200) and (220) peaks, forming a set of lines. The intercept of each line in the ordinate axis was used to calculate the value of $L$ [see Ref. 27 (Noheda et al.) for more details of calculation]. The coherence length decreases upon cooling, more sharply closer to the cubic to rhombohedral phase transition, then suddenly becomes temperature-independent at $T_C \approx 230$ K, and remains almost constant below $T_C$.

Figure 3 give the variation of the cubic or rhombohedral lattice constant, $a$, and the deviation of angle, $(90 - \alpha)$, as a function of temperature for PMN-xPT, together with the data on PMN taken from Ref. 31 for comparison. As in the case of PMN-5PT, the variation of the lattice constant $a$ of PMN-10PT and PMN-15PT shows an anomaly at $T_C \approx 275$ and 320 K, respectively, where the rhombohedral distortion appears with the angle α deviating from 90°. With increasing PT content, the following phenomena are observed: i) The lattice parameter *a* decreases; ii) The cubic/rhombohedral phase transition occurs at higher temperatures, from $T_C \approx 230$ K for PMN-5PT to $T_C \approx 320$ K for PMN - 15PT; iii) The rhombohedral distortion at low temperatures enhances, with α ≈ 89.95° for PMN-5PT, to α ≈ 89.875° for PMN-15PT, at 50 K; iv) The changes of lattice parameters at the cubic/rhombohedral transition become less sharp. Figure 4 gives the variation of the Curie temperature $T_C$ and the rhombohedral angle α (at 50 K) as a function of composition of PMN-xPT.

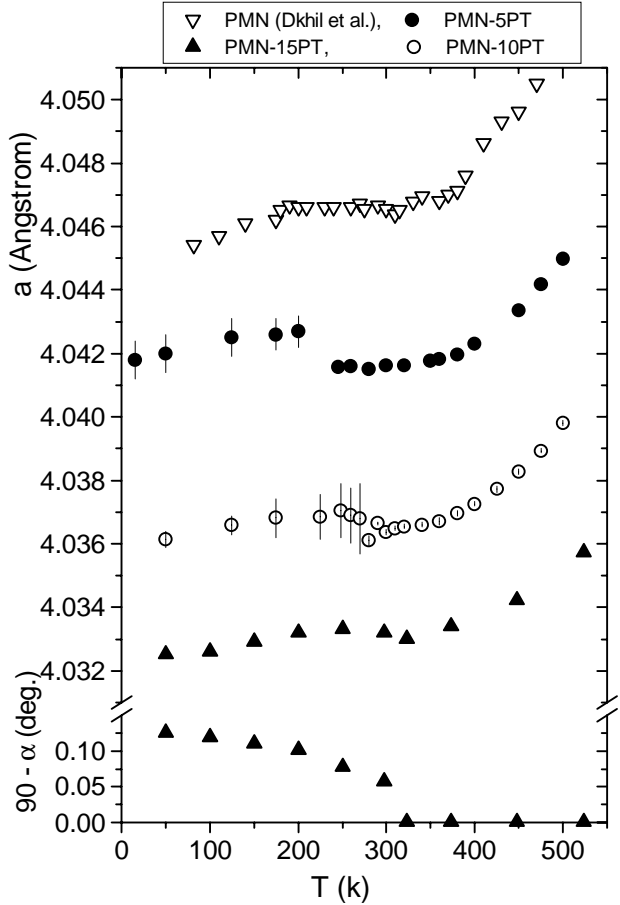

FIG. 3 Variation of the cubic and rhombohedral lattice constant, *a*, and the deviation of angle, (90 - α), as a function of temperature for (1-x)PMN-xPT (with x=0, 5, 10 and 15%). Data on pure PMN (x = 0) were taken from Ref. 31 (Dkhil *et al.*) for comparison.

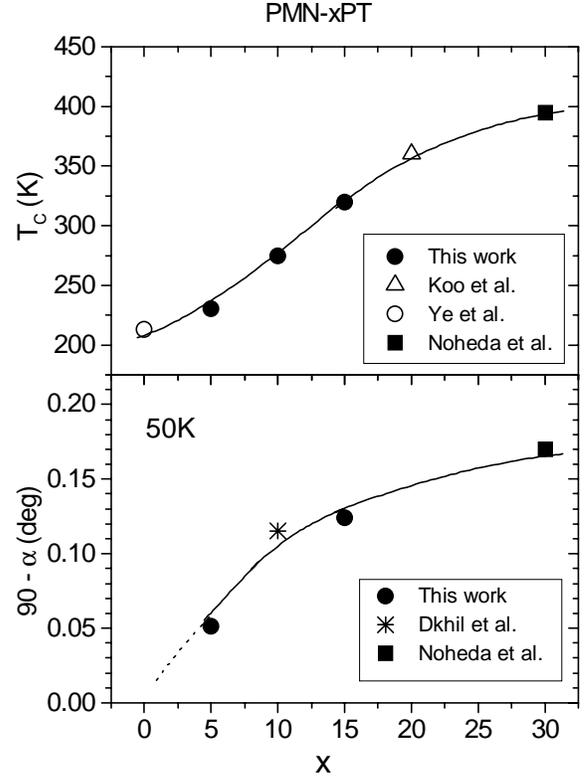

FIG. 4 Variation of the Curie temperature $T_C$ and the rhombohedral angle α (at 50 K) as a function of composition for PMN-xPT. Data on PMN ($T_C$), PMN-10PT (α), PMN-20PT ($T_C$), and PMN-30PT ($T_C$ and α) were taken from Ref. 16 (Ye *et al.*), Ref. 31 (Dkhil *et al.*), Ref. 32 (Koo *et al.*) and Ref. 27 (Noheda *et al.*), respectively, to better illustrate the trends.

## IV. DISCUSSION AND CONCLUSIONS

### A. Ferroelectric ordering in PMN-xPT

The structural analysis based on high-resolution synchrotron x-ray diffraction has allowed us to reveal some important aspects of the structural transformation in the solid solution system of PMN-xPT. Pure PMN exhibits an average cubic symmetry without significant split of the (hhh) peaks down to very low temperatures. The weak diffuse scattering observed at 50 K as a tail at the basis of the (222) reflection (Fig. 1) suggests, however, that local rhombohedral distortions indeed occur in PMN at low temperatures, probably with a short-range order. This observation is consistent with the previous work by de Mathan *et al.*[11] in which a two phase model, one of long-range average cubic symmetry, the other of rhombohedral polar symmetry due to atomic shifts giving rise to the local polar nanoregions, was introduced to better fit the static diffuse scattering observed in some (mainly high-angle) peaks of the x-ray and neutron diffraction patterns. Dkhil et al.[31] also confirmed that the polar order in PMN remains short-ranged.

On the other hand, the softening of the transverse optic lattice mode at the zone center in the paraelectric phase and the recovery of the overdamped soft mode below 220 K, recently observed by neutron scattering study,[18,19] together with the anomaly in the hypersonic damping,[20] clearly indicates the onset of ferroelectric order in PMN. The correlation length of this ferroelectric order, i.e. the dimensions of the PNRs in which the resulting structural distortion occurs in PMN, seem to be just near the detectable limit of the x-ray or neutron diffraction techniques, so that only weak diffuse scattering reflecting the local polar symmetry can be observed. The substitution of $Ti^{4+}$ for the complex B-site $(Mg_{1/3}Nb_{2/3})^{4+}$ ion is found to result in the decrease of lattice constant $a$ (obviously due to the smaller ionic radius of $R_{Ti4+}$ = 0.745 Å as compared with $R_{Nb5+}$ = 0.78 Å and $R_{Mg2+}$ = 1.16 Å [36]) and the development of a clean rhombohedral phase at a PT-concentration as low as 5%. In PMN-10PT and PMN-15PT, we have also observed a very sharp and clean rhombohedral split, indicating the full establishment of the polar phase. The phase transition into the rhombohedral phase is clearly seen as the abrupt change of lattice constant $a$ and the appearance of rhombohedral angle α (Figs. 2 & 3).

The jumps in $a$ and α at the transition become less pronounced with increasing $x$. This result suggests that the ferroelectric phase transition in PMN-PT with a low PT content is a first-order one, and it tends to be second-order with increasing Ti-amount. This behavior is consistent with the first-order nature of the phase transition from the field-induced ferroelectric state to the cubic phase observed in PMN at $T_C$ = 213 K[16] and the dielectric anomaly characteristic of second-order transition found in PMN-25PT.[37] Note that in the former case, the field-induced phase transition also tends to become 2$^{nd}$ order with increasing field strength (E) and a tricritical point exists in the T - E phase diagram [Ref. 1 (Ye)].

These observations provide some new insight into the development of the ferroelectric phase in PMN-PT with regards to the previously reported results that the average structure of PMN-PT remains to be pseudocubic at low PT content and the rhombohedral phase is established only at a PT-content higher than 13%,[29,30] even though a spontaneous rhombohedral to cubic phase transition was reported in PMN-10PT around 285 ºC, based on the temperature dependences of the dielectric constant[38] and the diffuse intensity and lattice parameter.[31]

At the same time, in the high-temperature phase of PNM-PT one can observe (Fig. 3) a strong deviation of $a(T)$ from the linear dependence that is a usual behavior for the paraelectric phase of the perovskite materials. The character of this deviation is qualitatively the same for all the compositions studied. This implies that the PNRs still exist in the high-temperature phase of PMN-PT at a significant concentration. As mentioned above, the PNRs can be considered as local regions of the rhombohedral phase that has a larger lattice dimensions $a$. Thus, the increase in the number and size of the PNRs with decreasing temperature starting from $T_d$ should lead to the progressive deviation of the average lattice constant from the linear trend expected in the pure paraelectric phase. In the same way, the decrease of the coherence length with decreasing temperature in the high-temperature phase [Fig. 2(b)] can be interpreted. The emergence of PNRs gives rise to the breaking of the coherency of the paraelectric phase due to the presence of the regions of a different (short-ranged rhombohedral) symmetry, as well as due to the internal strains introduced by these regions. At $T < T_C$, the coherence length, which is already that of the rhombohedral, but not the cubic, structure, does not change with temperature, indicating that the domain structure remains unchanged, the same as usually observed for a normal ferroelectric phase.

Thus, we have shown that the ferroelectric phase transition (i.e. abrupt change of structure and ferroelectric properties at certain temperature) is suppressed in PMN, but can be restored by addition of a small (≤ 5 %) amount of $PbTiO_3$. On the other hand, the relaxor-type behavior related to the formation and evolution of PNRs remains almost unchanged at $T > T_C$ even with the addition of 15 % $PbTiO_3$.

### B. Compositional disorder and polar nanoregions

The above results on the structural and dynamic properties raise, first of all, the following question: why the ferroelectric type mode-softening in PMN leads to the locally polar structural distortion, but not to the development of a long-range ferroelectric order. The general answer to this question is that the quenched compositional disorder in $Mg^{2+}$ and $Nb^{5+}$ ionic arrangement prevents the development of the long-range ferroelectric order. Two approaches have been proposed which are compatible with the idea of displacive-type ferroelectric phase transition disturbed by compositional disorder: i) the quenched spatial fluctuations of local Curie temperature leading to the formation of PNRs;[39] ii) the breaking of the ferroelectric long-range order into PNRs by quenched random fields.[40] But the specific mechanisms of the influence of the compositional disorder on the ferroelectric phase transition are still the topic of intensive discussion.

The "quenched random field" model [40] has been applied to relaxors based on an early idea of Imry and Ma [41] who showed that in a ferromagnetic-type system with a continuous symmetry of the order parameter, the second-order phase transition should be destroyed by an arbitrary weak static random field conjugate to the order parameter, i.e., below the Curie temperature the system becomes broken into small-size domains (analogy of PNRs) instead of forming a long-range ordered state as normally expected in the absence of the random field. The dimensions of the domains are determined by the interplay of the domain walls energy and the statistics of the random field. The same arguments are believed to be approximately applicable to PMN, in spite of the facts that the phase transition here is of first-order [16] and the symmetry of the order parameter (polarization) is not continuous. The disordered distribution of the heterovalent $Nb^{5+}$ and $Mg^{2+}$ ions on the same kind of lattice sites should provide the source of quenched random electric field.

To explain our results in the frame of this model, one may notice that the substitution of $Ti^{4+}$ for ($Mg^{2+}$/$Nb^{5+}$) ions decreases the random field effects, but it can hardly be considered as the main factor affecting the phase transition behavior since the long-range ferroelectric order can be destroyed by an *arbitrary weak* random field. Another factor that can really cancel the effects of random field is the transformation of the second-order phase transition into a first-order one (e.g. due to the change of composition). Such kind of transformation should manifest itself in the appearance of the jump of the order parameter at $T_C$ and thus in the increase of the domain wall energy just below $T_C$. The equilibrium number of domain walls in the crystal should become smaller in this case, or in other words, the dimensions of domains should become larger. But we have found the opposite behavior in PMN-PT, i.e., a growth of domains with increasing $x$ was observed alongside with the tendency for the phase transition to change from the first-order to the second-order. Therefore, we prefer to describe our results in terms of the approach implying the formation of PNRs as a result of the disorder-related fluctuations of the local Curie temperature, rather than as a result of the random field effects. Nevertheless, we do not necessarily deny the possible role of quenched random field. In particular, a strong random field can probably force the dipole moment of PNR to point parallel to the field, preventing reorientation of this moment due to the thermal motions or the interactions with other PNRs.

### C. Kinetics of phase transitions in relaxors

The next question to be answered is why the addition of a small amount of $PbTiO_3$ into PMN relaxor results in a normal ferroelectric phase that appears under a relatively sharp phase transition.

The relaxor to normal ferroelectric phase transition is often explained by the interactions between the PNRs, which are strong enough to provoke the ferroelectric-type ordering of the PNRs below the phase transition temperature. This approach implies that the relaxor to normal ferroelectric phase transition be considered as an order-disorder process, which is fundamentally different from the usual displacive-type ferroelectric phase transition related to the soft lattice mode. But such an ordering process alone does not result in an abrupt change of the average lattice constant $a$, as indeed observed at $T_C$ (Figs. 2 & 3).

To explain the relatively abrupt variation of the measured $a$ (or the average unit cell volume) at $T_C$, we assume a sharp increase of PNR dimensions (i.e. the change of the concentration of the rhombohedral phase, having a larger $a$, in the heterophase mixture), which can be naturally derived from the model proposed by Bokov.[42] This model does not consider the interactions between PNRs as the main factor leading to a ferroelectric order, even though it does not exclude the possible role of such interactions. Instead, it relates the behavior of phase transitions in disordered crystals to the transition kinetics. The approach used is analogous to the aforementioned treatment of random field instability [41] in that the balance of the bulk and the surface energy of domains (or PNRs) is believed to determine their dimensions, but it is also different from the latter in that a first-order phase transition is considered and the random field is not regarded as the main reason for the creation of PNRs. It was shown that in the course of cooling below $T_d$, the formation of the PNRs takes place in two stages. At the first stage (when the temperature is high), the clusters of the short-range ferroelectric order (i.e. PNRs) begin to appear in the regions with a higher local Curie point. Each of these clusters has an equilibrium size, which can grow with decreasing temperature. The number of PNRs also increases with decreasing temperature as more nanoregions transform into the polar state. When the temperature is close to the mean Curie temperature $T_C$ (denoted as $T_{0m}$ in Ref. 42), the second stage begins, namely, the thermally activated formation of critical ferroelectric nuclei from some PNRs and their (isothermal) growth. This process is analogous to the formation of a new phase at a normal first-order phase transition. The growth of a nucleus is restricted by the collisions with its neighboring growing nuclei (like in the case of a normal phase transition) or with the PNRs that cannot be reoriented and merged into the growing nucleus for some reasons (such as "freezing" of PNRs due to low temperature or *strong* local random fields).

The resulting phase behavior depends on the value of a parameter $\rho_c$ ($\propto T_C$), which determines the number of PNRs formed at the first stage. If $\rho_c$ is large, there are few polar clusters. As a result, at the second isothermal stage, the nuclei of the ferroelectric phase can grow almost unrestrictedly, leading to macro- or mesoscopic ferroelectric domains and lattice distortions that can be detected by x-ray or neutron diffraction. This seems to be the case for PMNT - PT with a significant concentration of PT, where $\rho_c$ is comparatively larger due to a higher $T_C$ resulting from the presence of the ferroelectrically active $Ti^{4+}$ ion on the disordered B-site [with increasing amount of PT, the ferroelectric phase is further enhanced with $T_C$ rising from 213 K in PMN to 320 K in PMN-15PT (Fig. 4)]. Moreover, with a higher $T_C$, many of the existing PNRs can more easily be reoriented and merged into the growing nuclei after their collision, additionally increasing the size of final domains. The jump in the $a(T)$ dependences (Figs. 2 & 3) should result from such a process, leading to the development of the rhombohedral symmetry in PMN-PT.

In PMN, however, $\rho_c$ is smaller (since $T_C$ is lower) and numerous PNRs already exist when the critical ferroelectric nuclei begin to grow. Besides, every PNR is the obstacle to the growth of nuclei because it is thermally "frozen". As a result, the second step of the process (i.e. the isothermal growth of nuclei) is suppressed and the anomaly in the $a(T)$ dependence is absent. The final size of the "domains" at low temperatures remains compatible to the size of the PNRs initially formed at high temperatures, thus too small to be detected by x-ray or neutron diffraction but large enough to give rise to the recovery of phonon soft mode and the anomaly in hypersonic damping.

Therefore, it is sensible to state that the difference between the nanopolar state in PMN and the rhombohedral phase in PMN-PT lies in the number and the size scale of the polar regions and their kinetics. The locally polar state in the relaxor PMN is therefore of ferroelectric nature and the rhombohedral phase detected in PMN-xPT (x $\geq$ 0.05) presents the natural development of the ferroelectric ordering triggered by Ti-substitution.

As another key element of our conclusions, we note that the ferroelectric order is established in PMN-xPT below $T_C$ because of the divergence of the critical scattering and the observed rhombohedral distortion. This picture seems to conflict with the neutron scattering observation that the strong diffuse scattering subsists below $T_C$ in all relaxors so far examined, including PMN-20PT.[32] We are trying to resolve this issue experimentally very soon. We also intend to investigate more extensively the (1-x)PMN-xPT system by a combination of structural, dielectric and optical studies in order to provide some deeper insight into such fascinating issues as i) the microscopic mechanism underlying the development of ferroelectric ordering through enhanced polar interactions, and ii) the relationship between the ferroelectric ordering / structural transition (at $T_C$), the relaxor behavior and the dynamics of the polarization regimes.

## ACKNOWLEDGMENTS

This work was supported by the Natural Science and Research Council of Canada (NSERC), the U.S. ONR Grant No. N00014-99-1-0738 and the U.S. DOE under Contract No. DE-AC02-98CH10886.